\documentclass[12pt,preprint]{aastex}

\usepackage{epsfig}
\usepackage{amsmath}
\usepackage{amssymb}
\usepackage{graphicx}
\usepackage{appendix}
\usepackage{natbib}
\usepackage{apjfonts}

\newcommand{\lSect}[1]{\label{sec:#1}}
\newcommand{\lFig}[1]{\label{fig:#1}}
\newcommand{\lEq}[1]{\label{eq:#1}}
\newcommand{\lTab}[1]{\label{tab:#1}}
\newcommand{\Msun}{\mbox{M$_\odot$}}

\newcommand{\iso}[2]{\ensuremath{^{#2}\rm{#1}}}

\def\gtaprx {\lower .1ex\hbox{\rlap{\raise .6ex\hbox{\hskip .3ex
	{\ifmmode{\scriptscriptstyle >}\else
		{$\scriptscriptstyle >$}\fi}}}
	\kern -.4ex{\ifmmode{\scriptscriptstyle \sim}\else
		{$\scriptscriptstyle\sim$}\fi}}}
\def\ltaprx {\lower .1ex\hbox{\rlap{\raise .6ex\hbox{\hskip .3ex
	{\ifmmode{\scriptscriptstyle <}\else
		{$\scriptscriptstyle <$}\fi}}}
	\kern -.4ex{\ifmmode{\scriptscriptstyle \sim}\else
		{$\scriptscriptstyle\sim$}\fi}}}
\newcommand{\FIGFF}[2]{\ref{fig:#2}{#1}}

\newcommand{\FIG}[2]{Fig.~\FIGFF{#1}{#2}}
\newcommand{\Fig}[1]{\FIG{}{#1}}

\newcommand{\Eqref}[1]{\ref{eq:#1}}
\newcommand{\Eqff}[1]{(\Eqref{#1})}

\newcommand{\Eq}[1]{eq.~\Eqff{#1}}

\newcommand{\Tab}[1]{Table \ref{tab:#1}}

\shorttitle{Nucleosynthetic Constraints on Supernova Masses}

\begin{document}

\title{Nucleosynthetic Constraints on the Mass of the Heaviest
  Supernovae}

\author{Justin M. Brown\altaffilmark{1} and S. E. Woosley\altaffilmark{1}}

\altaffiltext{1}{Department of Astronomy and Astrophysics, University
  of California, Santa Cruz, CA 95064; jumbrown@ucsc.edu}



\begin{abstract}
Assuming a Salpeter initial mass function and taking the solar
abundances as a representative sample, we explore the sensitivity of
nucleosynthesis in massive stars to the truncation of supernova
explosions above a certain mass. It is assumed that stars of all
masses contribute to nucleosynthesis by their pre-explosive winds, but
above a certain limiting main sequence mass, the presupernova star
becomes a black hole and ejects nothing more. The solar abundances
from oxygen to atomic mass 90 are fit quite well assuming no cut-off
at all, i.e., by assuming all stars up to 120 \Msun \ make successful
supernovae.  Little degradation in the fit occurs if the upper limit
is reduced to 25 \Msun. The limit can be further reduced, but the
required event rate of supernovae in the remaining range rises rapidly
to compensate for the lost nucleosynthesis of the more massive
stars. The nucleosynthesis of the $s$-process declines precipitously
and the production of species made in the winds, e.g., carbon, becomes
unacceptably large compared with elements made in the explosion, e.g.,
silicon and oxygen. However, by varying uncertain physics, especially
the mass loss rate for massive stars and the rate for the
$^{22}$Ne($\alpha,$n)$^{25}$Mg reaction rate, acceptable
nucleosynthesis might still be achieved with a cutoff as low as 18
\Msun. This would require a supernova frequency three times greater
than the fiducial value obtained when all stars explode in order to
produce the required $^{16}$O. The nucleosynthesis of $^{60}$Fe and
$^{26}$Al is also examined.
\end{abstract}


\maketitle

\section{Introduction}
\lSect{intro}

Just which massive stars explode as supernovae and which collapse to
black holes has been a topic of great interest for a long time. In
particular, stellar nucleosynthesis can be used to constrain the
maximum mass of the supernova that needs to (or can) explode in order
to explain the abundances of the elements seen in the Sun and
elsewhere \citep[e.g.][]{Twa82,Twa87,Mae92}. Previous works have
generally focused on the production of helium and oxygen and the ratio
$\Delta Y/\Delta Z$ where Y is the helium mass fraction and Z is the
heavy element fraction. \citet{Mae92}, for example, concludes that the
observed abundances are best fit if stars of all masses contribute
their pre-explosive winds, but only stars below 20 -- 25 \Msun
\ explode as supernovae (though see \citet{Pra94} who contests this
conclusion). The remainder presumably end as black holes which accrete
the remaining star, including all its heavy elements.

Added interest in this issue has been generated recently by the
increasingly tight observational constraints placed upon the masses of
presupernova stars. \citet{Sma09} finds no evidence for supernova
progenitors with masses over 20 \Msun. \citet{Hor11} also finds an
inconsistency with the measured rate of core collapse supernovae and
the cosmic star formation rate in the sense that more stars seem to
form than are observed to die as supernovae by a factor of about 2. Of
course, these arguments are not yet absolute. More massive supernovae
may be hidden in dust and the connection between main sequence mass
and presupernova mass relies on theory. Star formation rates and
supernova rates are not precisely known, but these constraints may
become tighter with time and certainly suggest that not all massive
stars make luminous supernovae.

On the theoretical front, it has been known for a long time that more
massive stars are harder to blow up than lower mass ones
\citep[e.g.][]{Fry99,Fry04}. With higher mass, the entropies around
the collapsing iron cores are greater, and the fall off of density
with radius is consequently more gradual. During the collapse this
means a greater accretion rate on the proto-neutron star that is
harder to reverse. \citet{Oco11} recently quantified this effect with
a ``compactness'' parameter (see their Fig. 9) and measured the
difficulty of blowing up stars in a 1D code as a function of that
parameter. Their conclusion, interestingly again was that stars
heavier than 20 \Msun \ were harder to explode.

These considerations motivate a revisit of the problem of stellar
nucleosynthesis as a function of mass. Using a fiducial model set of
solar metallicity models from \citet{Woo07}, the same model set used
by \citet{Oco11}, we study the nucleosynthesis and remnant masses
resulting if the supernova explosions are truncated above a certain
mass, $M_{\rm BH}$. A Salpeter initial mass function is assumed
\citep{Salpeter1955} with a slope of $\Gamma = -1.35$. We determine
not only the bulk nucleosynthetic properties like oxygen and remnant
mass as a function of $M_{\rm BH}$, but also examine the synthesis of
the $s$-process, the individual ratios of important intermediate mass
and light elements, and the synthesis of interstellar radioactivities,
$^{26}$Al and $^{60}$Fe.

\section{Models}
\lSect{model}

The yield tables of \citet{Woo07} give the nucleosynthesis of all
species from hydrogen through lead for supernovae resulting from
non-rotating massive stars with the following initial masses: 12--33
(every integer mass), 35--60 (every five masses), 60--80 (every 10
masses), 100, and 120 solar masses. The authors calculated explosions
for four sets of models parameterized by the mass cut and explosion
energy. Here we use their standard set for which the explosion energy
was $1.2 \times 10^{51}$ erg and the mass cut was located at the
``entropy jump'' where $S/N_A k$ = 4.0. These are their ``A'' models
and are the same models for which \citet{Zha08} \ calculated compact
remnant masses and \citet{Oco11} studied compactness. The spread of
these models can be seen in \Fig{salpeter}. Nucleosynthesis ejected by
the pre-explosive winds and in the explosions was archived separately
but is not available in \citet{Woo07} (see Figure 4 in their paper for
the mass loss prescription used). The values of some key species are
provided in an online supplement for this paper.  Using this grid of
nucleosynthetic yields, we constructed a stellar population using the
high-end initial mass function described by \citet{Reid2006} as
illustrated in \Fig{salpeter}.

\begin{equation}
\lEq{salpeter}
\Phi(M)\operatorname{d}(M)\propto M^{-2.35\pm0.1}\operatorname{d}(M)
\end{equation}
The total yields of the stellar population were calculated by
integrating the yields over the mass function, as described in
\Eq{integrate}, where $m_i$ is the total production (in solar masses)
of isotope $i$, and $E_i (M)$ and $W_i (M)$ are the total ejecta of
isotope $i$ from, respectively, the supernova explosion and the winds
in solar masses from a star with initial mass $M$:
\begin{equation}
\lEq{integrate} m_i=\int^{M_{upper}}_{12M_\odot}\Phi(M)E_i(M)\operatorname{d}M+\int^{120M_\odot}_{12M_\odot}\Phi(M)W_i(M)\operatorname{d}M
\end{equation}
We calculate the yields for the stable isotopes, varying the upper
mass limit to the values listed in \Tab{statistics}.

Our results are expressed in terms of a simple ``production factor''
for each isotope defined as
\begin{equation}
P_i=\frac{m_i/\sum_i{m_i}}{S_i},
\end{equation}
where $S_i$ is the mass fraction of the isotope in the Sun
\citep{Lodders2003}. The importance of the production factor
lies in its relationship to the supernova rate. It can be shown that
the production factor from high-mass progenitors is inversely
proportional to the number of supernovae estimated by the model,
assuming that the majority of the isotope's production comes from this
stellar population.

For comparison, we plot the production factors as a function of atomic
mass for a population with no upper bound in \Fig{120}
\citep[see also][]{Woo07}.

\section{Limits on $M_{\rm BH}$}
\lSect{discussion}

The mass of the heaviest supernova that has to explode is constrained
by a variety of observations. These include not only the
nucleosynthetic pattern of the intermediate mass elements, but
especially the light component of the $s$-process, the frequency of
supernovae, and the masses of compact remnants. The existence of a
cutoff mass also has interesting implications for $\Delta$Y/$\Delta$Z
and the synthesis of $^{26}$Al and $^{60}$Fe which we also briefly
discuss.


We choose to analyze the statistics of the isotopes lighter and
heavier than the iron group nuclei separately as we expect the
elements heavier than atomic mass 60 to be secondary nucleosynthesis
due primarily to the $s$-process or the reprocessing of $s$-process
isotopes in the original star. The mean and standard deviations of the
production factors of the lighter and heavier iron group nuclei are tabulated separately in
\Tab{statistics}.

\subsection{The Production of Carbon, Oxygen, and Intermediate Mass Elements}

When studying the nucleosynthesis of massive stars, theorists often
normalize to the production of $^{16}$O, as it is the third most
abundant isotope in the universe and comes almost entirely from
massive stars \citep{Langer1996}. For reference, we've included the
production factor of $^{16}$O in \Tab{statistics} as well.

In \Fig{low_iso} we compare the production factor of various elements
to that of $^{16}$O for variable $M_{\rm BH}$. For present purposes,
success is defined as being within a factor of two of the solar
abundances. We use a number of alpha elements to probe the strength of
the oxygen and silicon burning, and we use $^{14}$N to measure the
strength of the CNO cycle. From this, it's clear that little is lost
in terms of the production of these common isotopes if the upper mass
limit is reduced from 120 to 40 \Msun, which supports the preliminary
limits set by \citet{Heger2003}.

However, for low upper mass limits, $^{12}$C is overproduced in the
winds compared to $^{16}$O, suggesting either that this bound cannot
be much lower than 25 \Msun \ (a production factor of two) or that
there is significantly less mass loss than assumed. Carbon is mostly
made in the winds of very massive stars, especially during their
Wolf-Rayet stage.  Oxygen, which is the major part of the
metallicity is also made in winds but more in the explosions. Their
ratio then is quite sensitive to the mass loss prescription used. To
illustrate this, in \Fig{low_iso}, we've also included the ratios for
a population for which we've halved the mass loss yields. In the
figure, it's apparent that the C/O ratio remains consistent with the
solar abundances down to an upper mass limit of 20 \Msun. To
illustrate the change in the nucleosynthesis between the upper mass
limit extrema, we present \Fig{20} as a comparison to \Fig{120}. The
apparent large overproduction of $^{40}$K in both figures is not a
problem because $^{40}$K is radioactive with a half life of
$1.26\times10^9$ years \citep{crc2009}. Much of it will decay before
being incorporated into the Sun.

We have checked other important intermediate mass isotope ratios in
order to probe how well these stellar populations represent the solar
abundances. These include \iso{Si}{28}/\iso{Ca}{40},
\iso{Ca}{40}/\iso{Mg}{24}, \iso{Si}{28}/\iso{Fe}{56}, and
\iso{Ca}{40}/\iso{Fe}{56}. As can be seen in \Fig{newiso}, these all
remain acceptably close to solar ratios for the entire range of
$M_{BH}$.

\subsection{The Supernova Rate}

Using the production factor of $^{16}$O, we can estimate how the
supernova rate depends upon $M_{\rm BH}$. An accurate calculation of
the rate itself would require a galactic chemical evolution model that
includes gas accretion onto and outflows from the galactic disk---as
described in \citet{Tim95}---and is beyond the scope of this
paper. However, we can estimate the factor by which the supernova rate
would have to increase by examining $N_{SN}$, the total number of
massive stars required to die to produce the correct amount of
$^{16}$O, normalized to the total number required to produce the
yields of our control population (M$_{\rm BH}=120$ \Msun). The number
of stellar deaths (and therefore, supernovae) increases slowly with
the lowering of the upper mass limit until approximately 40 \Msun,
below which it increases rapidly up to twice the original number at
28 \Msun \ and three times the original number at 19 \Msun.

\subsection{The Light $s$-Process}

The heavier component of the $s$-process, those nuclei above A
$\approx$ 90 is thought to be produced in low mass stars
\citep{Pignatari2010}. The light component of the $s$-process, on the
other hand, is generally attributed to massive stars and occurs late
during helium burning when the temperature rises sufficiently for the
$^{22}$Ne($\alpha,$n)$^{25}$Mg reaction to occur.  Owing to the
temperature sensitivity of this rate, production of the $s$-process
isotopes in the mass range A = 60 -- 90 is a potentially powerful
constraint on M$_{\rm BH}$. It is important that we {\sl overproduce}
the $s$-process elements somewhat in this stellar mass range as many
of the primary isotopes are also made in lower metallicity stars that
don't produce many $s$-process elements.

Use of this diagnostic is complicated by the fact that many isotopes
in the mass range $A$ = 60 -- 90 can be produced by both the $s$ and
$r$-processes, and so an alternative approach is to focus on a few
``$s$-only'' isotopes (\Fig{high_iso})---those isotopes produced
exclusively by the s-process---$^{70}$Ge, $^{76}$Se, $^{86}$Sr, and
$^{87}$Sr \citep{Kappeler1989}. We exclude $^{80}$Kr and $^{82}$Kr
from our analysis---despite both being $s$-only isotopes---as these
isotopes are also significantly produced in low-mass asymptotic giant
branch stars. We notice that the $s$-process is most predominant in
core-collapse supernovae from 30 to 50 \Msun. With more complex models
of the stellar population, a comparison of the total $s$-process
yields to the solar abundance could prove to provide greater insight
as to the upper mass bound. Here, we see that $^{86}$Sr and $^{87}$Sr
are underproduced regardless of the upper bound, which could indicate
an issue with nuclear cross sections or our stellar models. The other
isotopes, $^{70}$Ge and $^{76}$Se appear to be produced in sufficient
quantities for us to use them for this paper. The production of
$^{70}$Ge falls below $1/2$ the solar abundance at 21 \Msun, whereas
the production of $^{87}$Sr falls below $1/2$ solar production at 23
\Msun. However, if we increase the rate of $^{22}$Ne($\alpha$,n) to
the maximum experimental value according to \citet{Jaeger:2001lr}, we
find that a limit of 18 \Msun \ is a reasonable value for the mass of
the heaviest supernovae.

\subsection{$^{60}$Fe and $^{26}$Al}
\lSect{fe60}

The nuclei $^{60}$Fe ($\tau_{1/2} = 2.62 \times 10^6$ y) and $^{26}$Al
($\tau_{1/2} = 7.17 \times 10^5$ y) are interesting because they
accumulate in the interstellar medium where the emission generated by
their decays can be studied using gamma-ray telescopes. Observations
by \citet{Smi04} give a ratio for the decay rate of $^{60}$Fe to that
of $^{26}$Al of about 0.16 implying a ratio of $^{60}$Fe to $^{26}$Al
mass fractions of (60/26)(0.16) = 0.37 \citep{Woo07}. As pointed out
by \citet{Pra04} and discussed in \citet{Woo07}, current calculations
give a larger value. Our current study uses yields that already
address some of the concerns discussed in \citet{Woo07}, including
inappropriate rates for $^{26}$Al destruction and use of OPAL
opacities for electron scattering at high temperature. As
\Fig{fe60al26} shows, our mass-averaged estimate of the production is
1.0 for $M_{\rm BH}$ = 120. This compares favorably with the value
0.95 given for these models by the authors.

The remaining difference between 1.0 and 0.37 probably chiefly
reflects remaining uncertainty in the neutron capture cross sections
for $^{59}$Fe and $^{26}$Al and especially the choice of explosion
energies for low mass supernovae. Studies of the light curves of Type
IIp supernovae \citep{Kas09}, suggest that many supernovae, perhaps
most, have an explosion energy smaller than the fiducial $1.2 \times
10^{51}$ erg assumed in many studies. Reducing the explosion energy
reduces the yield of $^{60}$Fe substantially. \citet{Lim06} have also
emphasized the dependence of this production ratio on mass loss,
convection theory, and the initial mass function. Including the
effects of rotation may also increase the $^{26}$Al yield
\citep{Pal05}, especially in very massive stars.

Our purpose here is not to completely solve the debate surrounding
$^{60}$Fe/$^{26}$Al but to point out that the answer depends upon
$M_{\rm BH}$. Measurements of gamma-ray line flux ratios may thus
ultimately help constrain the masses of stars that
explode. \Fig{fe60al26} shows that the ratio of $^{60}$Fe/$^{26}$Al
produced by supernovae declines as $M_{\rm BH}$ becomes smaller. Given
that the production of $^{60}$Fe will be even smaller in those stars
we are assuming still explode if their kinetic energy is reduced, this
effect could reduce the average production of $^{60}$Fe appreciably.

\subsection{Helium Production and $\Delta$Y/$\Delta$Z} 
\lSect{dydz}

As discussed extensively in \citet{Mae92}, the measured derivative of
the helium mass fraction with respect to metallicity can, in
principle, be used to constrain $M_{\rm BH}$.  This is because the
winds of the most massive stars are rich in helium while the heavy
elements are largely confined to the cores. If the cores collapse to
black holes, trapping the heavy elements, the synthesis of the winds
remains and increases the overall average of
$\Delta$Y/$\Delta$Z. Observations suggest a value $\Delta$Y/$\Delta$Z
$\sim$ 4.

In  practice,   the  application  of  this  metric   is  fraught  with
uncertainty. The  yields of the  massive stars are dependent  upon the
initial  mass function  and especially  the very  uncertain  mass loss
rates  employed. After  the helium  core  is uncovered  by mass  loss,
Wolf-Rayet mass  loss contributes not  only helium, but  an increasing
amount  of  carbon and  oxygen.  The  remnant  masses depend  upon  an
uncertain  explosion mechanism.  Does all  the presupernova  star fall
into the  hole or only part? For  successful explosions, where  is the
mass cut? Lower mass stars  also contribute appreciably to both helium
and metallicity and the yields of stars of all masses are sensitive to
metallicity.   Even   an   approximate  meaningful   result   requires
integration over some uncertain model for galactic chemical evolution.

However, because this metric has been applied extensively in the
literature, we give in \Fig{delydelz} our results for the solar metallicity
stars considered in this survey. Until $M_{\rm BH}$ is reduced below
$\sim$40 \Msun \ $\Delta$Y/$\Delta$Z remains slightly less than unity. Even for $M_{\rm BH}$ = 18 \Msun,
$\Delta$Y/$\Delta$Z is still only 1.929, well below the observed
value.

These results are, at first glance, seemingly inconsistent with those
of \citet{Tim95} who found $\Delta$Y/$\Delta$Z = 4 for $M_{\rm BH}$ =
17 \Msun \ for solar metallicity stars and 30 \Msun \ for low
metallicity stars. However, Timmes et al. used the survey of
\citet{WW95} which included only stars below 40 \Msun, while our grid
extends to 120 \Msun. Furthermore, Timmes et al. assumed that any
winds would be metal free and not extend into the helium core. We
include initial metals in the envelope and mass loss from Wolf-Rayet
stars once the core is uncovered. If we redo our calculation using the
assumptions of Timmes et al., the dashed curve in \Fig{delydelz}
results. This curve crosses $\Delta$Y/$\Delta$Z at $M_{\rm BH}$ = 17
\Msun, in excellent agreement with Timmes et al. However, we believe
that our results for solar metallicity are more realistic and that a
lower value of $\Delta$Y/$\Delta$Z from massive stars is appropriate.

\subsection{Average Mass of Compact Remnants}
\lSect{nstar}

We use Table 4 from \citet{Zha08} to get the baryonic masses of the remnants for our fiducial set of supernovae from stars of initial masses of 10 to 100 \Msun.
We then calculate the gravitational mass of the stellar remnants with their Equation 2. We assume that the maximum gravitational mass of a neutron star is 2.0 \Msun \, so all remnants with a higher gravitational mass than this form black holes with a gravitational mass equal to its baryonic mass. We then construct a population of stars according to a Salpeter initial mass function and vary the heaviest supernova mass. We let any star with $M>M_{\rm BH}$ become a black hole with mass equal to its presupernova mass. In \Fig{remnant}, we plot various quantities (the average remnant mass, the average black hole mass, the average neutron star mass, the average iron core mass, and the maximum remnant mass) as a function of $M_{\rm BH}$. For those stars more massive than the heaviest supernova, we have assumed that the presupernova star completelycollapses to a black hole, so our black hole masses are larger than those of \citet{Zha08}.

We can compare this to the observed average neutron star mass from \citet{Sch10}. From 14 neutron stars with well-measured masses, they find an average neutron star mass of $1.325\pm0.056$ \Msun. We find an average neutron mass closest to this at 14 \Msun \ and are within their uncertainty up to 16 \Msun. This suggests that a low value for $M_{\rm BH}$ is more consistent with the observed neutron star masses.

\section{Conclusions}
\lSect{conclusions}

By examining a variety of nucleosynthetic diagnostics, one can
constrain the mass of the maximum mass supernova that must explode and
not swallow up its heavy elements in a black hole. At the outset, a
number of caveats are worth stating. First, our yield set from
\citet{Woo07} reflects a particular choice of many uncertain
parameters---the treatment of convection, mass loss, key reaction
rates (like $^{12}$C($\alpha,\gamma)^{16}$O), and explosion
physics. We lump together both means of forming a black hole, direct
and by fall back and we ignore mixing. It is quite possible, probable
even, that some supernovae make black holes and yet still eject some
fraction of their presupernova nucleosynthesis. Mixing during the
fallback epoch can complicate the estimate of yields. This probably
affects the intermediate mass elements and iron more than helium,
carbon, oxygen and the $s$-process. Our models sample a single
metallicity---solar. While other models of different metallicity are
available, using them would require a more careful treatment of
galactic chemical evolution than is presently justified, given all the
other uncertainties. Metallicity will not greatly affect the supernova
yield of primary elements like oxygen and the intermediate mass
elements, but it does affect the mass loss. We attempted to partially
compensate for that by multiplying the integrated nucleosynthesis of
the wind by a constant factor of 0.5, as shown in \Fig{low_iso}. The
use of solar metallicity also overestimates somewhat the production of
secondary species such as the elements with odd nuclear charge and the
$s$-process \citep[e.g.,][]{Tim95}, so somewhat larger yields than
solar might be required than suggested by \Fig{20}. Finally, all our
models ignore rotation. Not only can rotation affect the mechanism and
symmetry of the explosion, but it also makes the mass of the helium
and carbon-oxygen core larger for a given main sequence. So our
derived mass limits may actually need to be smaller.

Given these limitations, the best we can say at the present time is
what supernova mass limits might be consistent with observations. The
idea of a limiting mass is itself an approximation, since the
compactness of the core is not a monotonic function of main sequence
mass \citep{Oco11}, especially in the interesting range 20 -- 35
\Msun. For still heavier stars, mass loss may shrink the helium core
so much that the presupernova helium core mass of say a 100 \Msun
\ star differs little from that of a 20 \Msun \ star. Such massive
stars are rare however, and their nucleosynthesis is mostly due to
their presupernova winds.

We have looked at several processes that limit $M_{\rm BH}$. As $M_{\rm BH}$ is decreased, the necessary rate of
``successful'' supernovae rises. For $M_{\rm BH}$ = 20 the rate is
2.88 times greater than for $M_{\rm BH}$ = 120.
Surprisingly, in reducing $M_{\rm BH}$ to 20 \Msun, the overall
nucleosynthesis of most isotopes from Ne to Ca with respect to
$^{16}$O is altered little and the production of some isotopes in the
iron group is actually improved. The greatest apparent problem is $^{22}$Ne; however, this can be
mitigated somewhat by reducing the amount of mass loss or by
increasing the $^{22}$Ne($\alpha$,n) rate.
Of the $s$-process only isotopes produced mainly in massive stars,
$^{70}$Ge becomes the limiting isotope for the mass of the heaviest
supernova, reaching half of solar value at 23 $\Msun$. This can be
extended to 18 $\Msun$ by increasing the $^{22}$Ne($\alpha$,n) rate.

In total, we find that we can easily reduce the mass of the heaviest
supernovae to 40 \Msun \ without any significant changes. For KEPLER's
standard values of nuclear rates and mass loss, there are only
moderate changes in these processes down to 25 \Msun. The limits
become increasingly severe for smaller masses. For $M_{\rm BH}$ = 25
\Msun, the stellar winds overproduce $^{12}$C with respect to $^{16}$O
by a factor of two, unless we reduce the mass loss in these stars by
two. At $M_{\rm BH}$ = 21 \Msun, the lighter elements are overproduced
in these more massive stars, and the $s$-process produces
only half the needed $^{70}$Ge unless the
$^{22}$Ne($\alpha$,n)$^{25}$Mg rate is increased. At 20 \Msun, even
with halved mass loss, the winds overproduce $^{12}$C by a factor of
two.

\section*{Acknowledgements}

We are grateful to Frank Timmes for very helpful correspondence
concerning the calculation of $\Delta$Y/$\Delta$Z.  This research has
been supported at UCSC by the National Science Foundation (AST
0909129) and the NASA Theory Program (NNX09AK36G). JB also received
support from the University of California Regent's Fellowship
Program. This research is supported in part by the Department of
Energy Office of Science Graduate Fellowship Program (DOE SCGF), made
possible in part by the American Recovery and Reinvestment Act of
2009, administered by ORISE-ORAU under contract no. DE-AC05-06OR23100.

\bibliography{bibliography}

\begin{figure}
\includegraphics[width=300pt,angle=270]{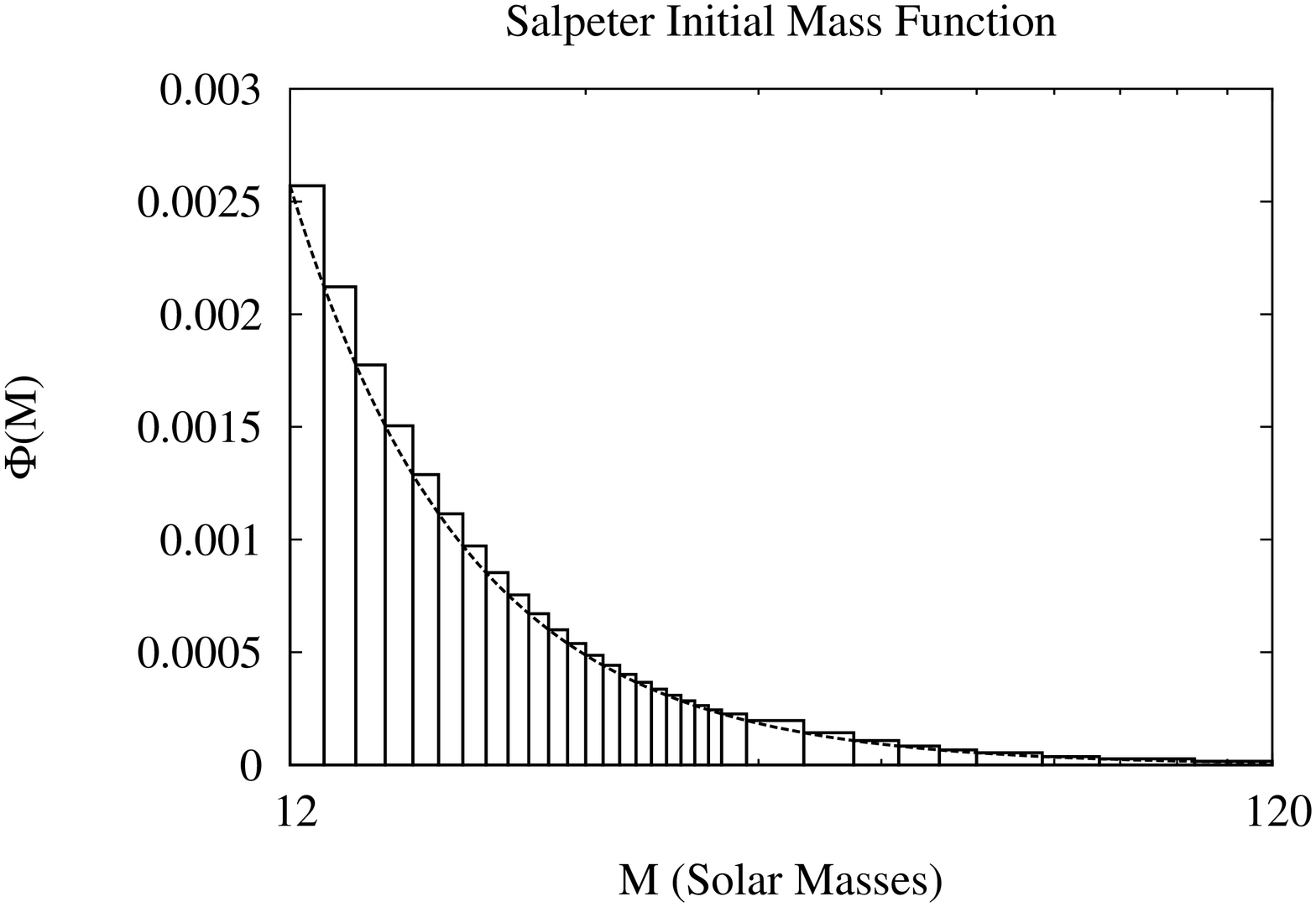}
\caption{\lFig{salpeter} The Salpeter Initial Mass Function is a model
  that represents the distribution of stellar mass in a
  population. Mass here is plotted logarithmically. The binning
  indicates the binning used in our grid of stellar models.}
\end{figure}

\begin{figure*}
\includegraphics[width=300pt,angle=270]{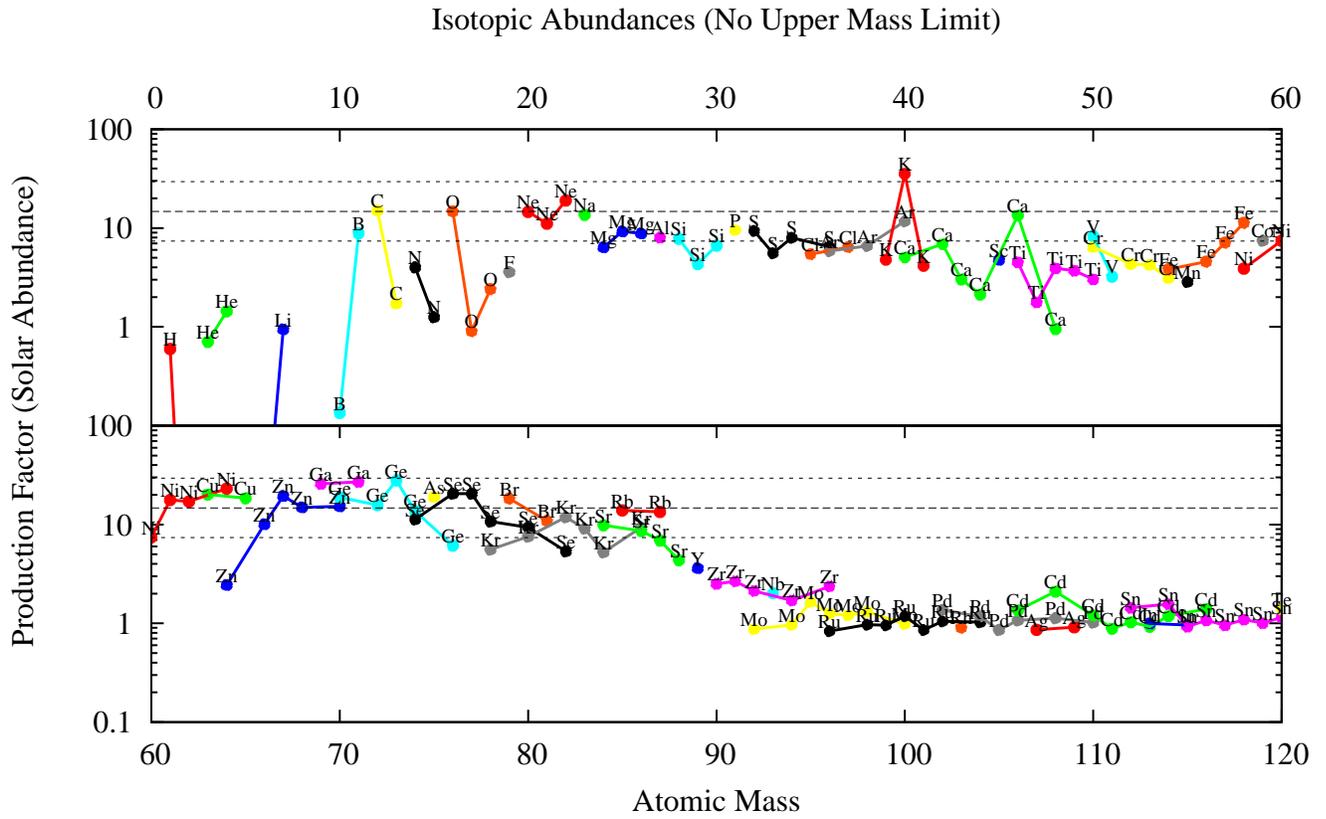}
\caption{\lFig{120} The production factor (plotted logarithmically) as
  a function of atomic mass. The dashed line represents the production
  factor of $^{16}$O. The dashed lines represent a factor of
  two deviation from $^{16}$O.}
\end{figure*}

\begin{figure}
\includegraphics[width=300pt,angle=270]{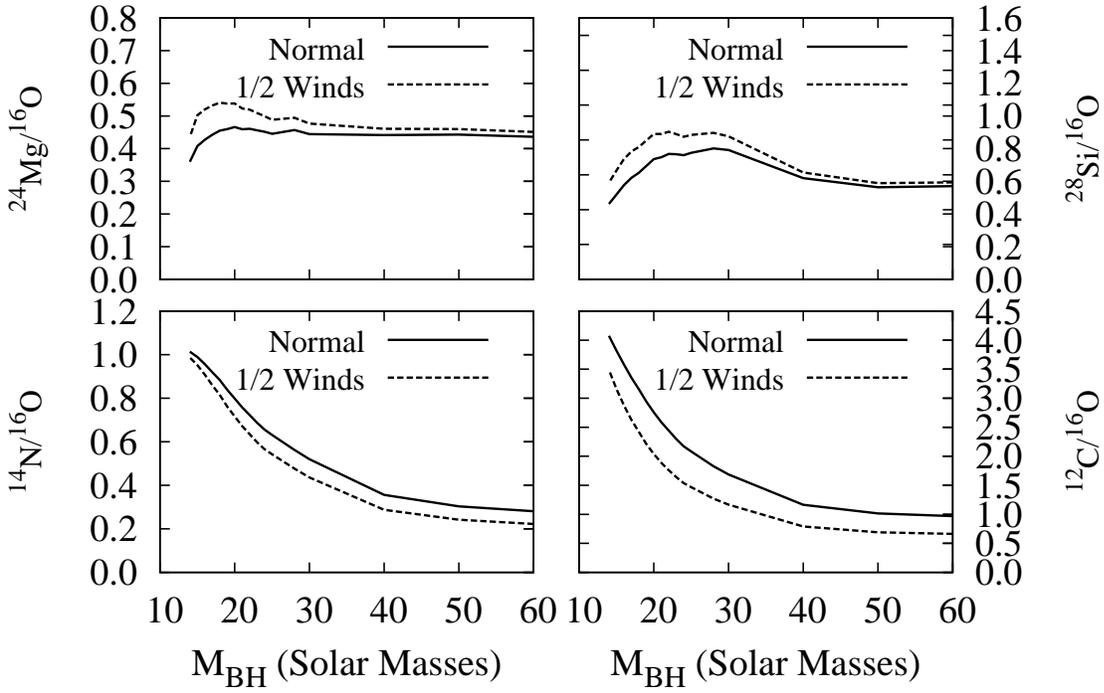}
\caption{\lFig{low_iso} The relative abundances of the common isotopes
  with respect to $^{16}$O as a function of $M_{\rm BH}$. The
  plot has been truncated at $M_{\rm BH}=60\Msun$, beyond which the
  abundances do not illustrate any significant variations. A relative
  abundance of 1 indicates solar ratio. The dotted lines represent the
  ratios produced if the wind contribution is halved to illustrate the
  dependence on wind.}
\end{figure}

\begin{figure}
\includegraphics[width=300pt,angle=270]{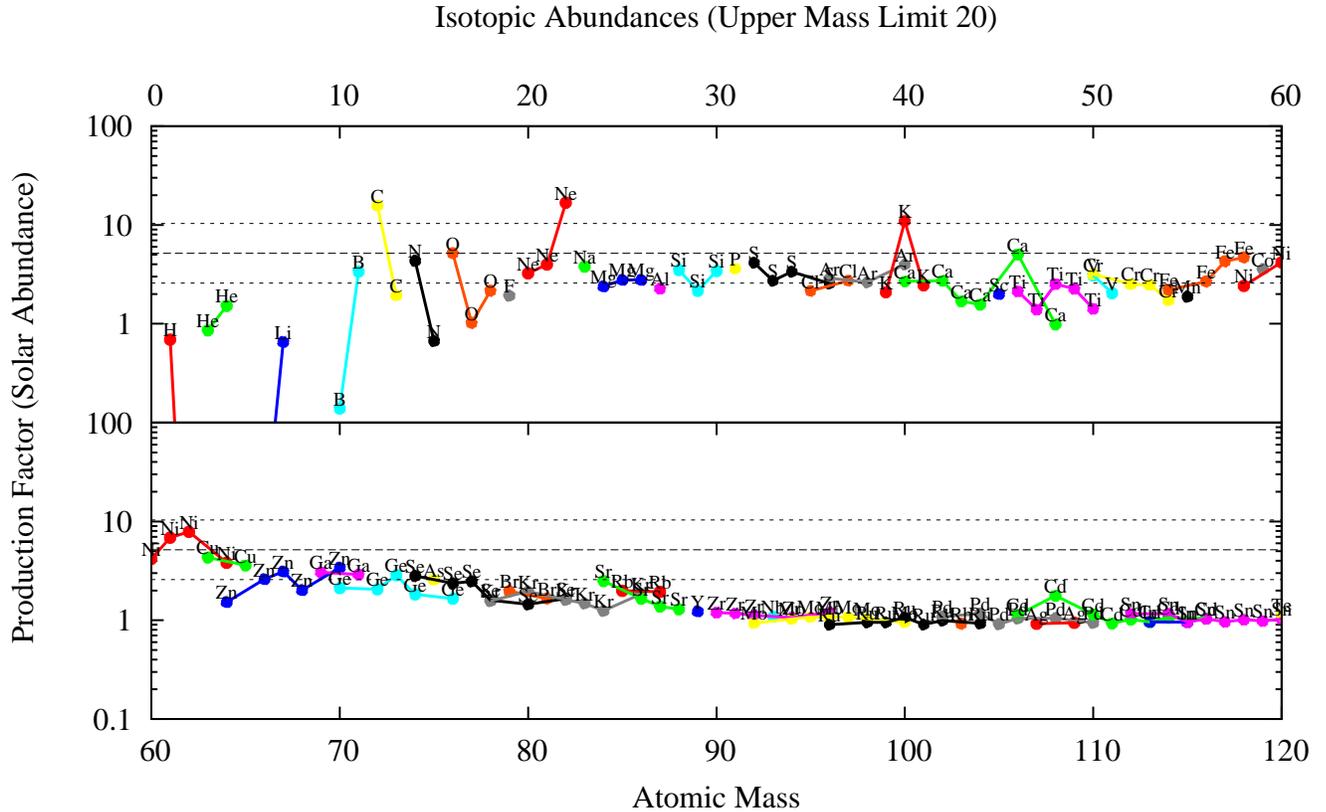}
\caption{\lFig{20} The production factor (plotted logarithmically) as
  a function of atomic mass. The dashed line represents the production
  factor of $^{16}$O. The dotted lines represent a factor of two
  deviation from $^{16}$O. Here it is assumed that all massive star
  over 20 \Msun \ end up collapsing as black holes and their only
  nucleosynthetic contribution is their winds. The large abundance of
  $^{40}$K is not a problem as this species will decay prior to solar
  system formation bu the large abundances of $^{22}$Ne and $^{12}$C
  could indicate unresolved issues with mass loss, nuclear physics, or
  the assumption of a low threshold for black hole formation. The
  $s$-process is also weak here (compare with \Fig{120}).}
\vspace{+1pt}
\end{figure}

\begin{figure}
	\includegraphics[width=300pt,angle=270]{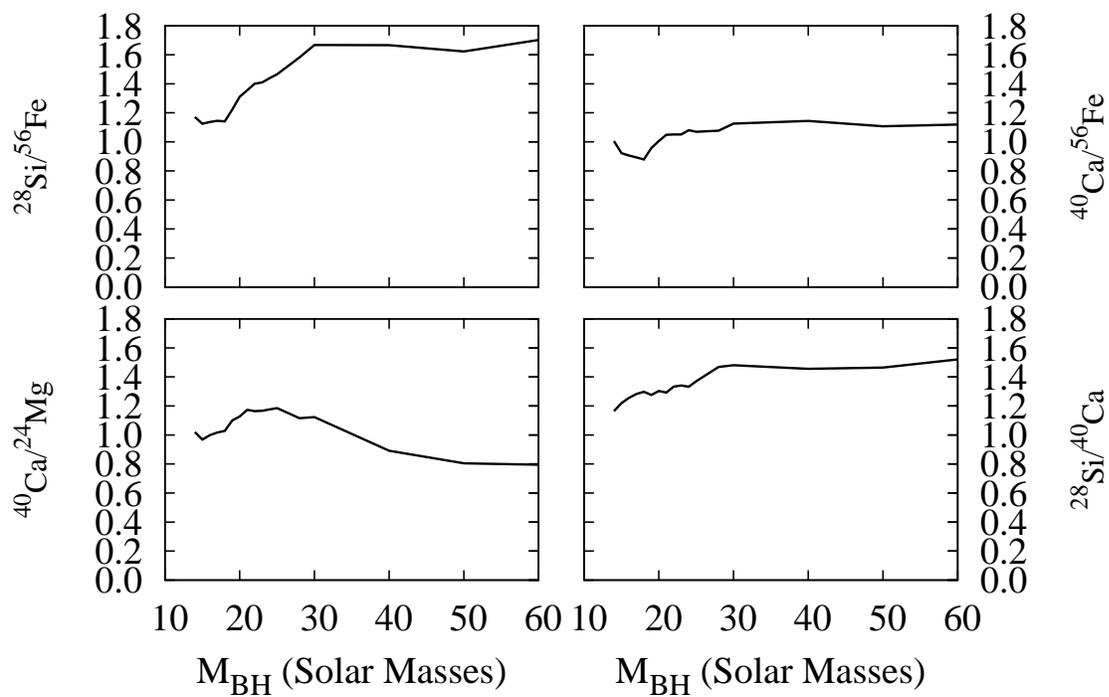}
	\caption{\lFig{newiso} The relative abundances of the
          intermediate mass isotopes as a function of $M_{\rm
            BH}$. The plot has been truncated at $M_{\rm BH}=60\Msun$,
          beyond which the abundances do not illustrate any
          significant variations.}
\end{figure}

\begin{figure}
\includegraphics[width=300pt,angle=270]{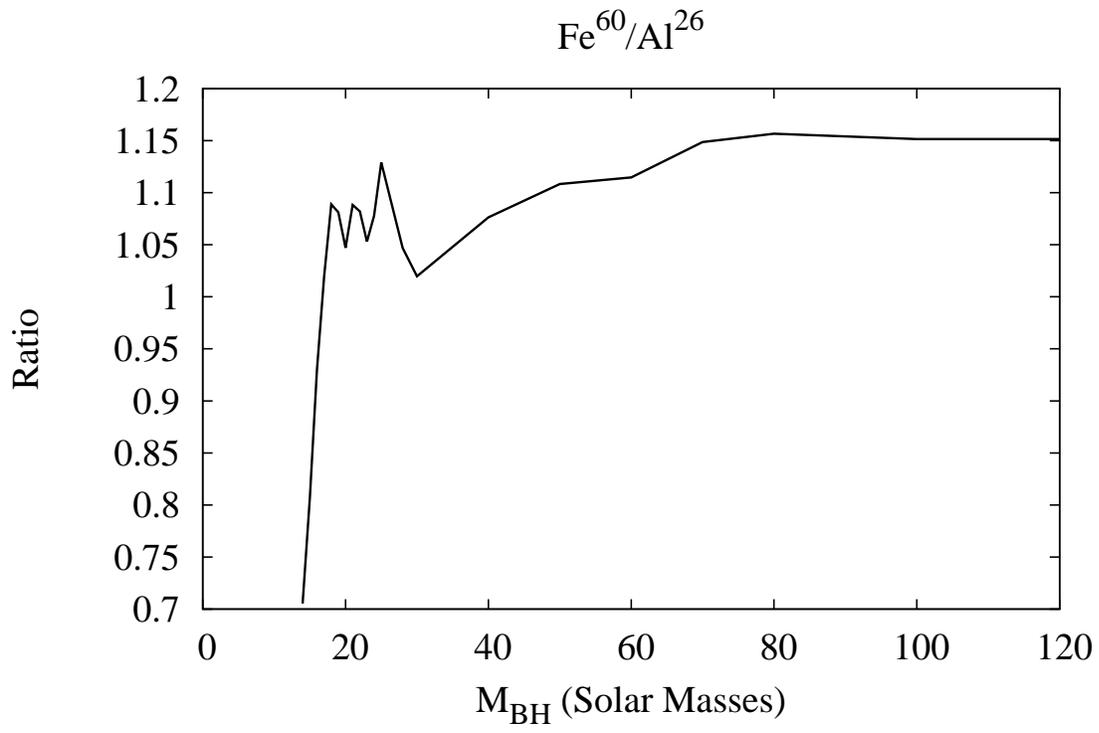}
\caption{\lFig{fe60al26} The ratio of the total masses of $^{60}$Fe to
  $^{26}$Al as a function of upper mass limit.}
\end{figure}

\begin{figure}
	\includegraphics[width=300pt,angle=270]{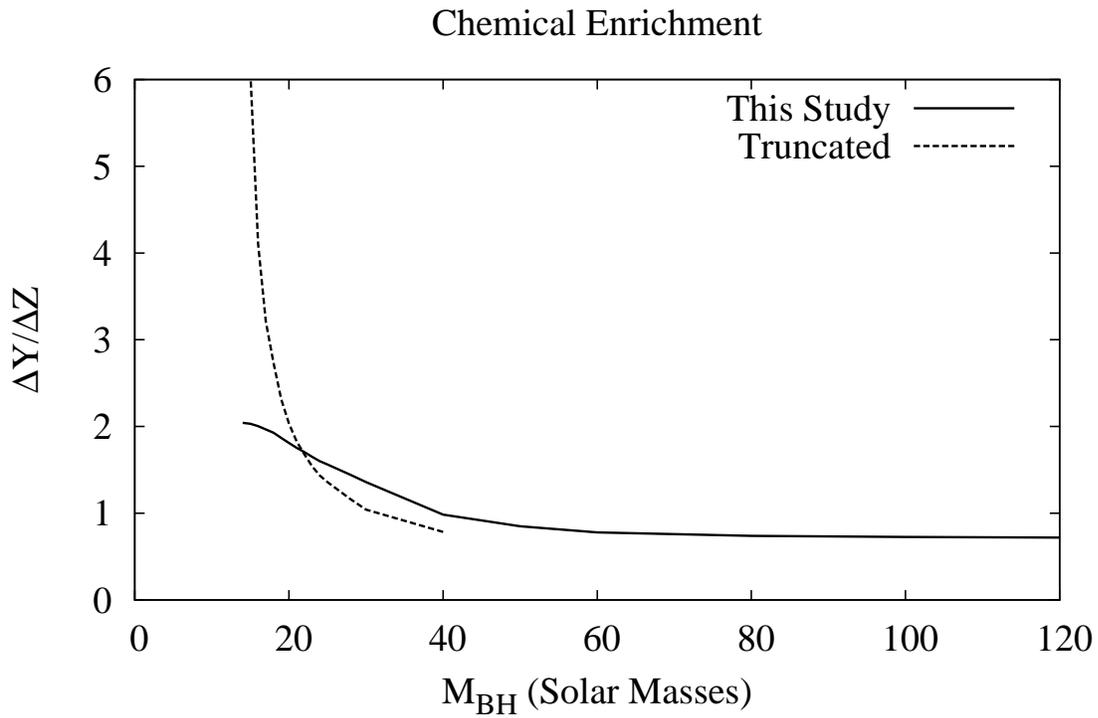}
	\caption{\lFig{delydelz} The ratio of the total deviations from the initial helium and metal mass fractions of the population as a function of the upper mass limit. The curve labeled ``Truncated'' uses the same assumptions as \citet{Tim95} in order to recreate the solar metallicity curve of Figure 37 of that same paper.}
\end{figure}

\begin{figure}
	\includegraphics[width=300pt,angle=270]{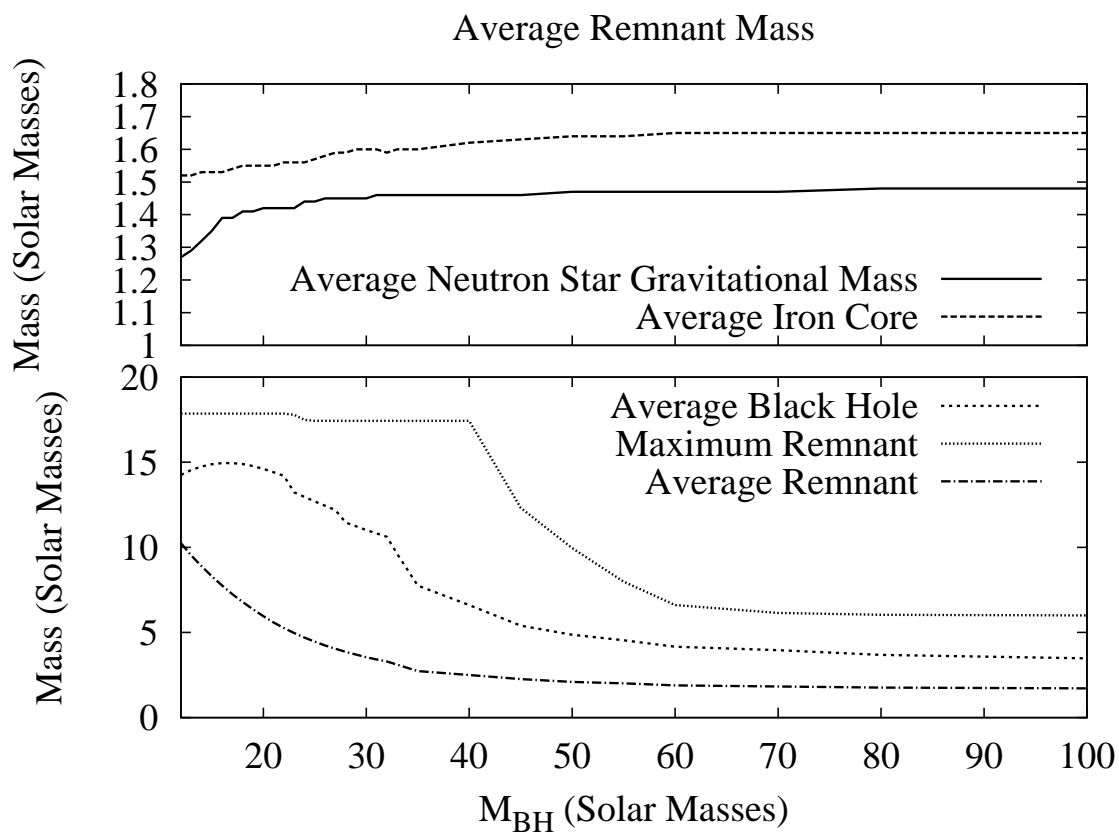}
	\caption{\lFig{remnant} The average neutron star remnant mass,
           iron core mass, average black hole mass, maximum remnant mass, and average remnant mass are plotted as a function of $M_{\rm
            BH}$. The average neutron star mass is within the
          observational uncertainty everywhere below $M_{\rm BH}=25$
          \Msun.}
\end{figure}

\begin{figure}
\includegraphics[width=300pt,angle=270]{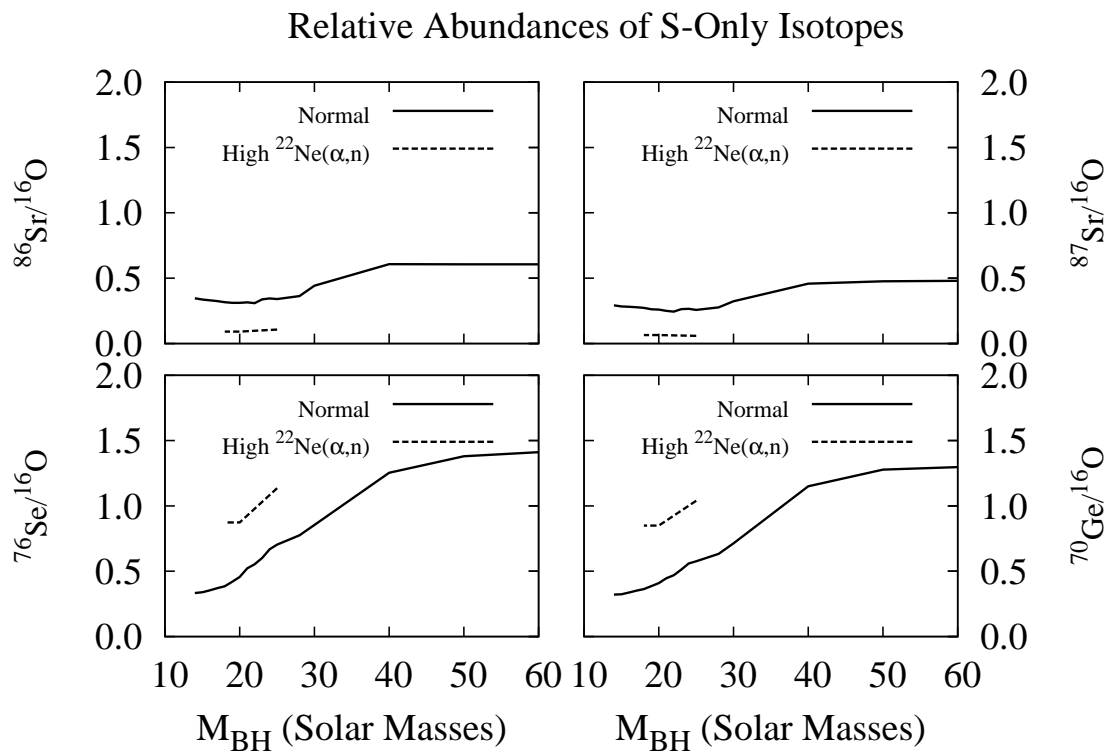}
\caption{\lFig{high_iso} The relative abundances of the $s$-only
  isotopes with respect to $^{16}$O and $^{24}$Mg as a
  function of $M_{upper}$. The plot has been truncated at
  $M_{upper}=60$\Msun, beyond which the abundances do not illustrate
  any significant variations. A relative abundance of 1 indicates
  solar ratio. The dashed lines represent the ratios produced if the
  $^{22}$Ne($\alpha$,n) is increased to the maximum experimental value.}
\end{figure}

\begin{deluxetable}{rllllll}
\tablecaption{\lTab{statistics} Production Factor Statistics}
\tablehead{\colhead{$M_{upper}$ (\Msun)} & \colhead{$^{16}$O}
  &  \colhead{Mean (16--59)} & \colhead{$\sigma$
    (16--59)} & \colhead{Mean (60--84)} & \colhead{$\sigma$ (60--84)}
  & \colhead{$N_{SN}$}}
\startdata
 120  &  14.74 & 6.21650 & 5.48443 & 14.59968 & 6.72117  &  1.0000000    \\
 100  &  14.63 & 6.18653 & 5.45782 & 14.54773 & 6.69476  &  1.0149254    \\
  80  &  14.48 & 6.14438 & 5.42532 & 14.48140 & 6.66672  &  1.0298507    \\
  70  &  14.24 & 6.04946 & 5.37006 & 14.34487 & 6.61504  &  1.0447761    \\
  60  &  14.01 & 5.96336 & 5.30620 & 14.16470 & 6.54938  &  1.0597014    \\
  50  &  13.14 & 5.60242 & 4.78830 & 13.20545 & 6.08493  &  1.1343282    \\
  40  &  11.46 & 5.19207 & 4.36852 & 10.63252 & 4.68389  &  1.2985075    \\
  30  &   7.93 & 4.23886 & 3.51987 & 5.86116 & 2.37390  &  1.8805970    \\
  28  &   7.33 & 3.97234 & 3.26334 & 5.26788 & 2.19142  &  2.0298507    \\
  25  &   6.52 & 3.60184 & 2.96335 & 4.35242 & 1.87591  &  2.2835820    \\
  24  &   6.27 & 3.49595 & 2.91884 & 3.95001 & 1.78152  &  2.3731341    \\
  23  &   5.99 & 3.37648 & 2.87740 & 3.59129 & 1.71603  &  2.4925373    \\
  22  &   5.69 & 3.25718 & 2.84823 & 3.29207 & 1.60152  &  2.6119401    \\
  21  &   5.44 & 3.14271 & 2.82677 & 3.01019 & 1.51820  &  2.7462685    \\
  20  &   5.17 & 3.00155 & 2.81088 & 2.69667 & 1.46384  &  2.8805971    \\
  19  &   4.93 & 2.82092 & 2.76516 & 2.51345 & 1.39076  &  3.0149255    \\
  18  &   4.70 & 2.66811 & 2.78619 & 2.30015 & 1.31519  &  3.1791043    \\
  17  &   4.52 & 2.55969 & 2.81806 & 2.11734 & 1.16281  &  3.2985075    \\
  16  &   4.36 & 2.43725 & 2.87535 & 1.94110 & 1.00387  &  3.4179103    \\
  15  &   4.22 & 2.31317 & 2.96168 & 1.78076 & 0.83437  &  3.5223880    \\
  14  &   4.12 & 2.20852 & 3.07487 & 1.60545 & 0.62216  &  3.6119401    \\
\enddata
\end{deluxetable}

\end{document}